\documentstyle[aps,twocolumn]{revtex}
\begin{document}

\title{ Corner overgrowth: Bending a high mobility two-dimensional
electron system by 90$^\circ$}

\author{
M. Grayson$^{(a)}$, D. Schuh, M. Huber, M. Bichler, and
G. Abstreiter}

\address{
Walter Schottky Institut, Technische Universit\"at M\"unchen, D-85748 
Garching, Germany\\}

\date{2 Nov 2004}

\maketitle
 
\begin{abstract}

Introducing an epitaxial growth technique called corner overgrowth, we 
fabricate a quantum confinement structure consisting of a high-mobility 
GaAs/AlGaAs heterojunction overgrown on top of an ex-situ cleaved substrate
corner.  The resulting corner-junction quantum-well heterostructure 
effectively bends a two-dimensional electron system (2DES) at an atomically 
sharp $90^{\rm o}$ angle.  The high-mobility 2DES demonstrates fractional 
quantum Hall effect on both facets.  Lossless edge-channel conduction
over the corner confirms 
a continuum of 2D electrons across the junction, consistent with 
Schroedinger-Poisson calculations of the electron distribution.  This growth
technique differs distinctly from cleaved-edge overgrowth and enables a 
complementary class of new embedded quantum heterostructures.

\end{abstract} 
\narrowtext

The fabrication of quantum confined electron systems with increasingly
complex geometries will become important as advances in technology reduce
the size and dimensionality of electronic devices.  In standard
low-dimensional semiconductor systems, the confinement potential is
atomically sharp only in the epitaxial growth direction and smooth on the
order of several $100$ \AA~ by gating or etching in the in-plane
directions.  One approach to achieve sharper in-plane confinement is to
imbed quantum heterostructures within a crystal by using a sequence of
orthogonal growths. Ground-breaking work using crystal regrowth techniques
\cite{Zaslavsky} and cleaved-edge overgrowth \cite{Pfeiffer} utilized {\em
two} distinct growth steps to produce high-mobility transport structures
with sharp quantizing potentials in multiple directions
\cite{Zaslavsky,Yacoby,Grayson,Chang,Kang,Majumdar,Deutschmann,Huber,Auslaender}.  
In this Letter we introduce a new corner overgrowth technique that, in a
{\em single} growth step, creates atomically sharp quantum confinement in
two perpendicular directions which meet at a corner junction.  We apply
this technique to fabricate an L-shaped junction between two orthogonal
high-mobility two-dimensional electron systems (2DES) and demonstrate
lossless flow of quantum Hall edge channels across the junction as proof
of the continuity of the 2D system.

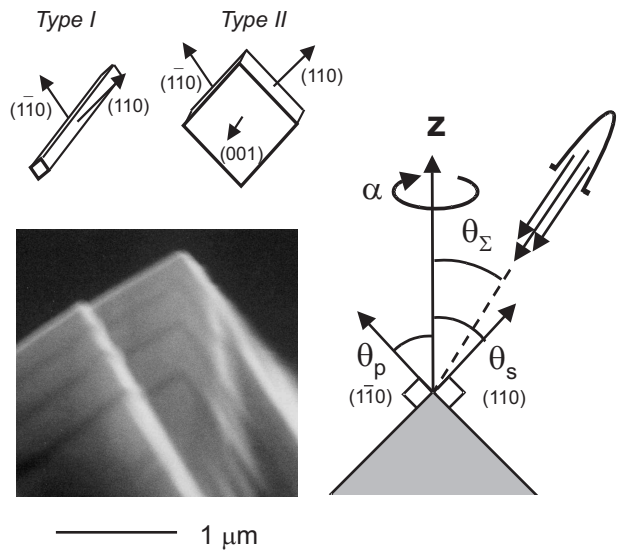
\begin{figure} 
\caption{
(Top) Cleaved corner substrates:  Type I from
(110) wafers, Type II from (001) wafers.  (Bottom) SEM micrograph
of corner overgrowth test structure. (Right) Schematic of growth
geometry: rotation axis $\hat{\bf z}$, source angle
$\theta_\Sigma$, substrate angle $\theta_s$, and precleave angle
$\theta_p$.
} 
\label{Fig1}
\end{figure}

We call our device the corner quantum-well heterojunction (CQW),
fabricated by overgrowing a standard GaAs/AlGaAs heterojunction structure
on an ex-situ cleaved corner.  To achieve high-quality growth across the
corner, both facets must have the same Miller index class.  Since 
ex-situ
cleaving is used to create the atomically sharp corner-substrate, we will
restrict growths to two particular geometries that expose orthogonal
\{110\}-class planes.  One can either cleave a (110)-GaAs substrate wafer
once in the orthogonal $(1\bar{1}0)$ plane (Fig. \ref{Fig1}, Type I), or one can
take a (001)-GaAs wafer and cleave twice in the (110) and $(1\bar{1}0)$
planes (Fig. \ref{Fig1}, Type II).  In both cases, the shared corner between the
(110) and $(1\bar{1}0)$ planes serves as the overgrowth corner and is
mounted facing the molecular flux.  For shorthand, we refer to (110) as
the $s$-facet and $(1\bar{1}0)$ as the $p$-facet (corresponding to
'substrate' and 'precleave', respectively, for Type I samples).

The samples were grown in a Epi Gen-II molecular beam epitaxy (MBE) system
with an ultra-high vacuum ambient ($P \approx 10^{-11}$~mbar).  The 
$\bf{\hat z}$-axis in Fig. \ref{Fig1} is the rotation axis of the substrate 
as well as the axis around which the molecular sources are symmetrically 
arranged at an azimuthal angle $\theta_\Sigma = 33^\circ$.  The molecular 
flux $\Phi_0$ is first calibrated with RHEED on a standard flat substrate. 
The corner-substrate is then mounted in its place with both $s$- and 
$p$-facets at a nominal $\theta_s = 90^\circ - \theta_p = 45^\circ$ angle.  
Under continuous rotation of angle $\alpha$ around the $\bf{\hat z}$ axis, 
the flux incident on the tilted $s$-facet will consist of an oscillating 
component which averages to zero, and a constant component equal to the 
RHEED calibrated flux $\Phi_0$ projected onto the tilted surface, $\Phi_s =
\Phi_0~cos(\theta_s)$ \cite{angle}.  The complementary $p$-facet will
correspondingly see an average flux of $\Phi_p = \Phi_0~cos(\theta_p) =
\Phi_0~sin(\theta_s)$.  One main result of this work is that in spite of
fluctuations in the molecular flux due to rotation, we nonetheless achieve
high quality crystal growth on the tilted facets of the corner substrate.  
After oxide desorbtion, the sample is overgrown under rotation
($d\alpha/dt$ = 7 rpm) at a substrate temperature of $T_{sub} =
460^\circ$C with all molecular fluxes increased by a factor of
$1/cos(45^{\rm o}) = \sqrt{2}$ to compensate for the geometrically reduced
flux on the tilted substrate.  For a typical growth, the beam equivalent
pressure BEP$_{\rm As_4} = 5 \times 10^{-5}$~mbar, and BEP$_{\rm Ga} = 2.2
\times 10^{-7}$~mbar, corresponding to a RHEED calibrated Ga growth rate
of $2.4$~\AA/s on a flat substrate.

In the SEM picture in Fig \ref{Fig1}. we confirm the sharpness of the corner 
growth
morphology in a test corner-overgrown superlattice structure.  Using the
RHEED calibrated flux values, the three dark bands on the left $p-$ (right
$s-$) side of the sample are 810~\AA ~ (560 \AA) AlAs separated by lighter
GaAs/AlGaAs layers of varying thickness 1620~\AA, 1620~\AA ~ and 2430~\AA
~ (1120~\AA, 1120~\AA, and 1680~\AA) from the surface downwards.  The
corner junction clearly maintains a sharp $90^\circ$ corner profile even
under regrowth.  The slightly thicker layers on the left result from a 
growth angle of $\theta_s = 56^\circ$ with the ratio of thicknesses $d$ on 
the two sides equal to $d_s/d_p = tan(\theta_s)$.  The jagged cleave is a 
consequence of forcing a break in a non-cleavage (001) plane,
perpendicular to the two natural cleave planes (110) and $(1\bar10)$.

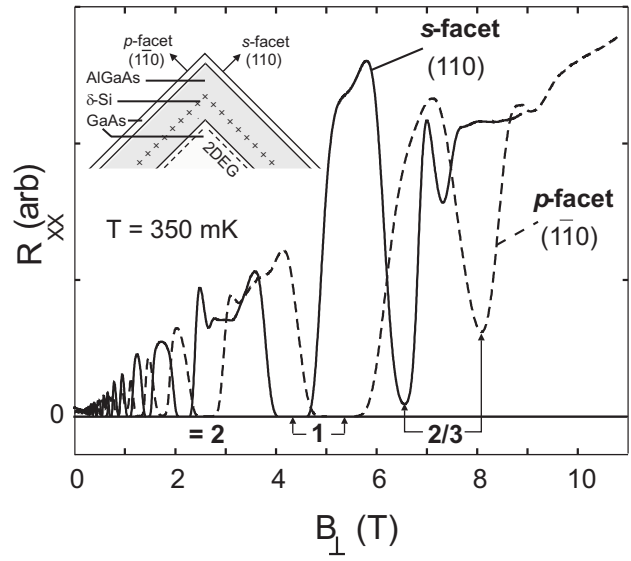
\begin{figure} 
\caption{
$R_{xx}$ for the two facets, with densities $n_s = 1.07 x 10^{11}$ 
cm$^{-2}$, $n_p = 1.30 x 10^{11}$ cm$^{-2}$.  Fractional QH minima 
at $\nu = 2/3$ attest to sample quality.  Inset:  Bent heterojunction 
quantum well structure.
} 
\label{Fig2}
\end{figure}

The transport structure that is the focus of this paper is a CQW
heterojunction, analogous to a single hetero-interface 2DEG in planar
structures.  The growth consists of an AlGaAs/GaAs superlattice buffer
and a base layer of GaAs, followed by growth on each facet of nominally
1200~\AA ~ Al$_{0.3}$Ga$_{0.7}$As, Si-$\delta$ doping, 3000\AA ~
Al$_{0.3}$Ga$_{0.7}$As, and a 100~\AA ~ GaAs cap layer (see Fig. \ref{Fig2} 
inset).  The sample is electrically contacted with indium, alloyed onto the 
sample edges away from the overgrown corner.  By measuring the 4-point 
longitudinal resistance of the two facets at 350 mK (Fig. \ref{Fig2}), the 
densities of the two facets can be independently measured with the result:  
$n_s = 1.07 \times 10^{11}~{\rm cm}^{-2}$ and $n_p = 1.30 \times 
10^{11}~{\rm cm}^{-2}$.  Observation of fractional quantum Hall effect 
minima forming at this strength at filling factor $\nu = 2/3$ at 350~mK 
attests that both facets have a transport mobility of order $\mu \approx 
5 \times 10^5$~cm$^2$/Vs.  

We explain the dissimilarity in the two densities by considering that the
thickness of growth on the two facets will not be equal if the corner is
not mounted at {\em exactly} $45^\circ$ during growth.  Assuming that the
dopant density is on the verge of parallel conduction at $\theta_s =
45^\circ$ as per design, the curve calculated in Fig. \ref{Fig3} shows the 2D
density $n_s(\theta_s)=n_p(90^\circ-\theta_s)$ as well as the parallel 
conduction density of the $\delta$-doping layer $n_\delta$ as a function 
of substrate tilt angle $\theta_s$. One sees the predominant effect of 
large tilt angles ($\theta_s > 45^\circ$) is thinner layers, and a 
corresponding rapid depletion of the 2D density due to the proximity of the 
surface pinning potential.  Conversely, with smaller tilt angles ($\theta_s 
< 45^\circ$) the layers are thicker, and the density drops only slightly 
due to the increasing spacer thickness $d$ between the 2D and the saturated 
dopant layer.  With decreasing $\theta_s$, the excess doping rapidly 
accumulates electron density in the parallel conducting dopant layer.  A 
growth angle of $\theta_s = 47^\circ$ would account for the appropriate 
ratio of $n_s/n_p$ observed in the measured device.

To demonstrate continuity of the 2D electron system, we first measure
the 2-point resistance across the corner at zero $B$.  The cross-corner
resistance is $\sim 1$ k$\Omega$ comparable to the 2-point resistivities
within a given facet, our first indication of a continuous 2D system.
For a more rigorous demonstration of 2D continuity, we apply a magnetic 
field and tilt the sample to an angle where the filling factors on the 
two facets make simple integer ratios.  Fig. \ref{Fig4} shows the case 
$\nu_s/\nu_p=2 ~(\theta_s= 31.3^{\rm o})$, and the inset shows the edge-state 
diagram for the $\nu_s:\nu_p = 2:1$ case.  With a current supplied from 
contact Y to contact Z the 4-point resistance across the corner between 
voltage contacts A and B, $R^{cc}_{AB}$, is zero ($0 \pm 0.5 \Omega$) 
whenever both 2D systems are gapped (dashed line, $\nu_s:\nu_p$ = 2:1, 4:2, 
6:3).  From the Landauer-Buttiker diagram, this means the outermost edge 
channel is transmitted across the corner with no backscattering, 
demonstrating the continuity of the 2D electron system across the corner. 
For voltage contacts C-D in the same gapped regions, the 
Landauer-Buttiker equations predict the observed $R^{cc}_{CD} = 
\frac{\nu_s - \nu_p} {\nu_s \nu_p}\frac{h}{e^2}$.

We calculated the self-consistent Hartree electron density
at the CQW heterojunction, revealing an enhanced density exactly at the
corner.  Integrating the electron density cross-sectionally in the (100)
direction and projecting onto the corner profile indicated by the dashed
line in the inset of Fig. \ref{Fig3}, we arrive at a measure of the local 2D density
along the CQW, unfolded and plotted as a solid line in Fig. \ref{Fig3}, 
inset.  We observe a single, deeply bound 1D-wire ground state at the 
corner $(d=0)$ with calculated ground energy $E_F - E_0 = 7$~meV and 
spatial extent $\Delta x_0 = 200$ \AA (dotted line).  All other states 
have much weaker binding energy $E_F - E_{n\ge1} \le 4$~meV and much 
larger spatial extent $\Delta x_{n\ge1} > 1000$ \AA, so we subtract out 
the 1D density and label the remainder as two-dimensional in character 
(dashed line).  Breaking the total charge density into 2D and 1D 
densities, we observe in the Fig. \ref{Fig2} inset, that the 2D states 
maintain a remarkably uniform density $n = 1.2 \times 
10^{11}~{\rm cm}^{-2}$ up to within $\lambda_F/2$ of the corner, where
$\lambda_F/2 = \sqrt{\pi/2n} = 360$~\AA ~ is half the Fermi wavelength.  
We also see small Friedel oscillations away from the corner with the
expected period $\lambda_F/2$.  We note that the calculated 2D
continuum of electrons across the corner is consistent with the observed
low cross-corner resistance and lossless edge channel conduction.

In conclusion, we have created a new kind of quantum confinement structure
by overgrowing an ex-situ cleaved corner substrate.  Self-consistent
Hartree calculations predict a tightly bound wire state at
the junction, along with a continuum of 2D electrons which approach the
corner with a remarkably uniform density.  The bent 2D electron system
shows high mobility and continuity as characterized with magnetotransport.

This work was supported in part by the DFG
Schwerpunktprogramm Quanten-Hall-Systeme (SFB 348).  M.G. thanks the A.v.
Humboldt Foundation for support, J. Smet and K. von Klitzing for
stimulating discussion, and S. F. Roth for SEM pictures.

$(a)$ Email: matthew.grayson@wsi.tum.de

\begin{figure} 
\caption{
Calculated 2D $s$-facet carrier density $n_s$ and parallel
conduction carrier density in $\delta-$dopant layer $n_\delta$ plotted as a
function of substrate growth angle $\theta_s ~ (\theta_p = 90^\circ -
\theta_s)$.  Inset: Hartree calculation of charge density (solid), 
separated into 1D (dotted) and 2D (dashed) contributions.
} 
\label{Fig3}
\end{figure}

\begin{figure} 
\caption{
4-point cross-corner resistances. $R^{cc}_{AB}=0$ and 
$R^{cc}_{CD} = \frac{\nu_s-\nu_p}{\nu_s \nu_p}\frac{h}{e^2}$ when both 
facets are gapped, as predicted by the Landauer-Buttiker formalism.
(inset). 
} 
\label{Fig4}
\end{figure}


\begin{references}
\bibitem{angle}
This simplified picture is valid as long as $\theta_s, \theta_p < 
90^\circ - \theta_\Sigma$.
\end{references}
\end{document}